\documentclass[a4paper,prd,preprint,aps,dvips,showpacs]{jpconf}
\usepackage{graphicx}
\usepackage[english]{babel}
\usepackage{amscd}
\usepackage{epsfig}
\usepackage{tabularx}
\usepackage{latexsym}
\usepackage{amsmath}
\usepackage{amsfonts}
\usepackage{amssymb}
\usepackage[latin1]{inputenc}
\usepackage{epsfig}
\usepackage{times}
\usepackage[T1]{fontenc}
\usepackage{graphics}
\usepackage{verbatim}
\usepackage[absolute]{textpos}
\usepackage{wrapfig,times}
\usepackage{amsthm}
\usepackage{setspace}

\newcommand{\be}{\begin{equation}}
\newcommand{\ee}{ \end{equation}}
\newcommand{\ben}{\begin{eqnarray}}
\newcommand{\een}{\end{eqnarray}}

\begin{document}
\title{Application of Fibonacci oscillators in the Debye model}

\author{Andr\'e A.A. Marinho$^{a,b}$, F.A. Brito$^{b}$, C. Chesman$^{a}$}

\address{$^a$ Departamento de Física Teórica e Experimental, Universidade Federal do Rio Grande do Norte,
Caixa Postal 1634, 59072-970 Natal, Rio Grande do Norte, Brazil 
\\
$^b$  Departamento de Física, Universidade Federal de
Campina Grande, 58109-970 Campina Grande, Paraiba, Brazil}

\ead{andre@dfte.ufrn.br or andreofisico@hotmail.com.br}

\begin{abstract}
In this paper we study the thermodynamics of a crystalline solid by applying $q$-deformed algebra of Fibonacci oscillators 
through the generalized Fibonacci sequence of two real and independent deformation parameters $q_1$ and $q_2$.
We find a  $(q_1, q_2)$-deformed Hamiltonian and consequently the $q$-deformed thermodynamic quantities.
The results led us to interpret the deformation parameters acting as disturbance or impurities factors modifying the 
characteristics of a crystal structure. More specifically,  we found the 
possibility of adjusting the Fibonacci oscillators to describe the change of thermal conductivity 
of a given element 
as one inserts impurites.
\end{abstract}


\section{Introduction}

The study of quantum groups and quantum algebras has attracted great interest in recent years and stimulated intense research 
in various fields of physics \cite{bie1,mac,anat} taking into account a range of applications covering astrophysics  
and condensed matter, for instance, black holes and high-temperature superconductors \cite{wil}. 

A possible mechanism to generate a deformed version of the classical statistical mechanics consists in replacing the Gibbs-Boltzmann 
distribution by a deformed version. In this sense it is postulated a form of deformed entropy which implies a 
generalized theory of thermodynamics \cite{tsa1, kan2, abe, lav9}. In this context, it was demonstrated that a natural realization 
of the $q$-deformed thermodynamics can be done via $q$-calculation formalism \cite{lav, liu}.

A new proposal for the $q$-calculation is the inclusion of two distinct deformation parameters in some physical applications. 
Starting from the generalization of $q$-algebra \cite{jac1}, in Ref.~\cite{arik2} was generalized the Fibonacci sequence. 
Here, the numbers are in that sequence of generalized Fibonacci oscillators, where new parameters ($q_1$, $q_2$) are introduced \cite{aba11,amg1,bri4}.
They provide a unification of quantum oscillators with quantum groups, keeping the degeneration property of the spectrum invariant under the symmetries of the quantum group. 
The quantum algebra with two deformation parameters may have a greater flexibility when it 
comes to applications in realistic phenomenological physical models \cite{dao,gong}.

We know that a solid is formed by a large number of atoms bound by cohesion forces of various types. 
The motion of atoms in a solid is very narrow, causing each atom to move only within a small neighborhood, executing 
vibratory motion around its equilibrium point. In a crystalline solid, the equilibrium points of atomic vibrations form a regular 
spatial structure, a cubic structure, for example. 

The study conducted by Anderson, Lee and Elliot \cite{ande, lee, ell} shows that the presence of defects or impurities in a 
crystal modifies the electrostatic potential in their neighborhood, breaking the translational symmetry of the periodic potential. 
This perturbation can produce electronic wave functions located near the impurity, ceasing to be propagated throughout the crystal.

The conductivity of semiconductors can be dramatically altered by the presence of impurities, i.e., different from atoms that make up the pure crystal. 
This property allows the production of a variety of electronic devices of the same 
semiconductor material. This process of placing impurities in semiconductor materials is called doping.

In the present study we follow the lines of  \cite{bri1,bri2,bri3} to address the issues of ($q_1, q_2$)-deformed thermodynamics through application of a generalized $q$-algebra to a Debye solid.
We find that the deformation is related to phenomena due to impurities or disorder factors in the system. 
We shall mainly show that the Fibonacci oscillators may act as defects or impurities in the crystal lattice, allowing 
us to modify quantities such as Debye temperature, thermal and electrical conductivities.

\section{Algebra of the Fibonacci oscillators}
The generalization of integers usually is given by a sequence. The two well-known ways to describe a sequence 
are the arithmetic and geometric progressions. However,  the Fibonacci sequence encompasses both. 
Generalizing this sequence, we get the Fibonacci oscillators, so the spectrum can now be given by the 
Fibonacci integer.

Defining the Fibonacci \textit{basic number}, the $q$-deformed quantum oscillator is now defined by the Heisenberg algebra 
in terms of the annihilation and creation operators $c$ and $c^{\dagger}$, respectively, and the number operator  $N$, as follows:

\be \label{e49}[x_{i,q_1,q_2}] = c_{i}^\dagger c_{i} = \frac{q_2^{2x_i}-q_1^{2x_i}}{q_2^2-q_1^{2}},\ee
\be c_i c_{i}^\dagger - q_1^{2}c_{i}^\dagger c_i = q_2^{2N_i},\qquad\mbox{and}\qquad c_i c_{i}^\dagger - 
q_2^{2}c_{i}^\dagger c_i = q_1^{2N_i},\ee
\be [1+x_{i,q_1,q_2}] = q_1^{2}[x_{i,q_1,q_2}]+q_2^{2N_{i}},\quad\mbox{or}\quad\
[1+x_{i,q_1,q_2}] = q_2^{2}[x_{i,q_1,q_2}]+q_1^{2N_i}.\ee

Let us now generalize to ($q_1, q_2$)-deformed algebra the study we accomplished in \cite{bri2,bri3} for only one $q$-deformation parameter.
Now we have the following Hamiltonian with the respective energy eigenvalues
\ben \label{e11.5}{\cal H}_{q_1,q_2} = \frac{\hbar\omega}{2}\left(cc^\dagger + c^\dagger c\right),\een
\be \label{eq2}E_{n_{i,q_1,q_2}} = \frac{\hbar\omega_E}{2}\Big([n_{i,q_1,q_2}]+[n_{i,q_1,q_2}+1]\Big)=
\frac{\hbar\omega_E}{2}+\frac{\hbar\omega_E\left(2\ln(q_2)-2\ln(q_1)\right)n}{q_2^2-q_1^2}.\ee
Notice that when $q_1=1$ and $q_2\to 1$ (and vice-versa) the deformation vanishes, so that we have the usual definition
\be E_n = \frac{\hbar\omega_E}{2}\left(2n+1\right).\ee

\section{ ($q_1,q_2$)-deformed Debye Solid }

Corrections of Einstein's model are given by the Debye model, allowing us to integrate from a continuous spectrum of
frequencies up to the Debye frequency $\omega_D$, as the number of normal modes of vibration is 
$3N$, as is found in the literature \cite{patt,hua,kit2}. Following the same mathematical development, and inserting the
given parameters ($q_1,q_2$), we can write the specific heat for any temperature in the form:
\begin{equation} c_{V_{q_1,q_2}}(T) = 3\kappa_{B}D(\alpha_{0_{q_1,q_2}}), \end{equation}
where $D(\alpha_{0_{q_1,q_2}})$ is the (${q_1,q_2}$)-deformed Debye function, defined by
\begin{equation} \label{eq12} D(\alpha_{0_{q_1,q_2}}) = \frac{3}{(\alpha_{0_{q_1,q_2}})^3} \int_{0}^{\alpha_{0_{q_1,q_2}}} 
\frac{\alpha^4\exp{(\alpha)}}{[\exp(\alpha)-1]^2}d\alpha, \end{equation}
\ben\alpha_{0_{q_1,q_2}} = \frac{\hbar\omega_{D_{q_1,q_2}}}{\kappa_{B}T} =  \frac{\theta_{D_{q_1,q_2}}}{T},\qquad 
\omega_{D_{q_1,q_2}} = 2\omega_D\left(\frac{\ln(q_2)-\ln(q_1)}{q_2^2-q_1^2}\right). \een 
where $\omega_{D_{q_1,q_2}}$ and $\theta_{D_{q_1,q_2}}$, are the $(q_1, q_2)$-deformed Debye frequency and temperature, 
respectively, and $\omega_D$ is the Debye frequency characteristic. Integrating Eq.(\ref{eq12}) by parts one finds 
\ben \label{eq12.5} D(\alpha_{0_{q_1,q_2}})&=&-\frac{3{\alpha_{0_{q_1,q_2}}}}{\exp({\alpha_{0_{q_1,q_2}}})-1}+\frac{12}
{\alpha_{0_{q_1,q_2}}^3}\Bigg\{-\frac{\pi^4}{15}-\frac{\alpha_{0_{q_1,q_2}}^4}{4}+\alpha_{0_{q_1,q_2}}^3
\ln[1-\exp(\alpha_{0_{q_1,q_2}})]\nonumber\\
&+&3\alpha_{0_{q_1,q_2}}^2Li_2[\exp(\alpha_{0_{q_1,q_2}})]-6\alpha_{0_{q_1,q_2}}Li_3[\exp(\alpha_{0_{q_1,q_2}})]+6Li_4
[\exp(\alpha_{0_{q_1,q_2}})]\Bigg\},\een
where 
\be Li_n(z)=\displaystyle\sum_{k=0}^{\infty}{\frac{z^n}{k^{n}}},\ee 
is the polylogarithm function. We can check that, as in the usual Debye solid, the low-temperature specific heat in a q-deformed Debye 
solid is proportional to $T^3$, due to phonon excitation, a fact that is in agreement with experiments. 
Thus, let us express the ($q_1,q_2$)-deformed specific heat for low temperatures as follows: 
\begin{eqnarray} c_{V_{q_1,q_2}} = \frac{12\pi^4\kappa_B}{5}\left(\frac{T}{\theta_{D_{q_1,q_2}}}\right)^3 = 1944\left(\frac{T} 
{\theta_{D_{q_1,q_2}}}\right)^3 \frac{J}{mol K}. \end{eqnarray}

Through the settings for the thermal conductivity $(\kappa)$ and electrical $(\sigma)$ --- see \cite{zim}, 
we can rewrite the $(q_1,q_2)$-deformed specific heat in the useful form
\begin{eqnarray}\kappa_{q_1,q_2}=\frac{\kappa c_{V_{q_1,q_2}}}{c_{V}} \qquad\qquad \mbox{and} \qquad\qquad 
\sigma_{q_1,q_2}=\frac{\kappa_{{q_1,q_2}}\sigma}{\kappa}.\end{eqnarray}

\begin{figure}[htb!]
\centerline{
\includegraphics[{angle=90,height=7.0cm,angle=270,width=7.0cm}]{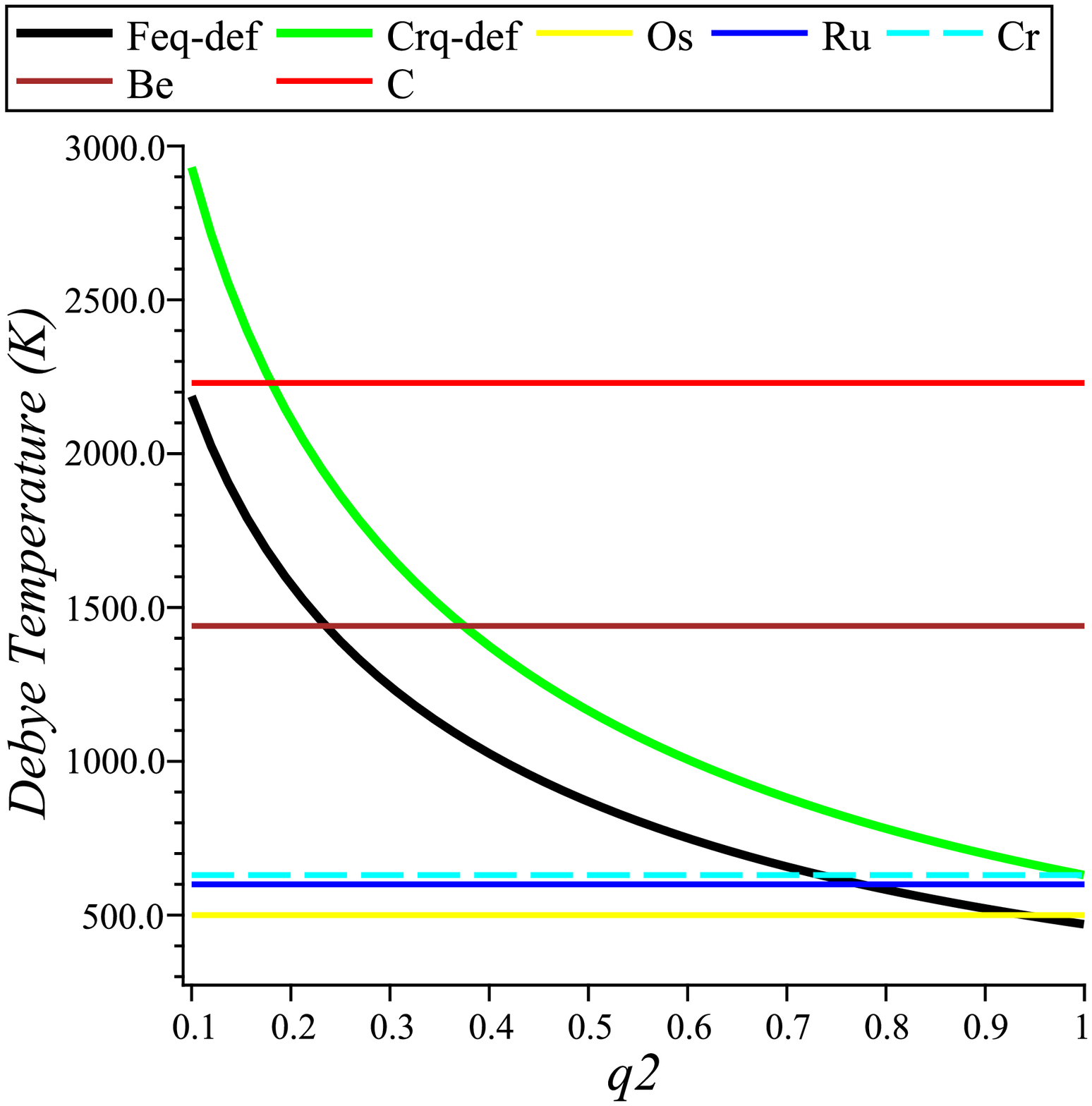}
\includegraphics[{angle=90,height=7.0cm,angle=270,width=7.0cm}]{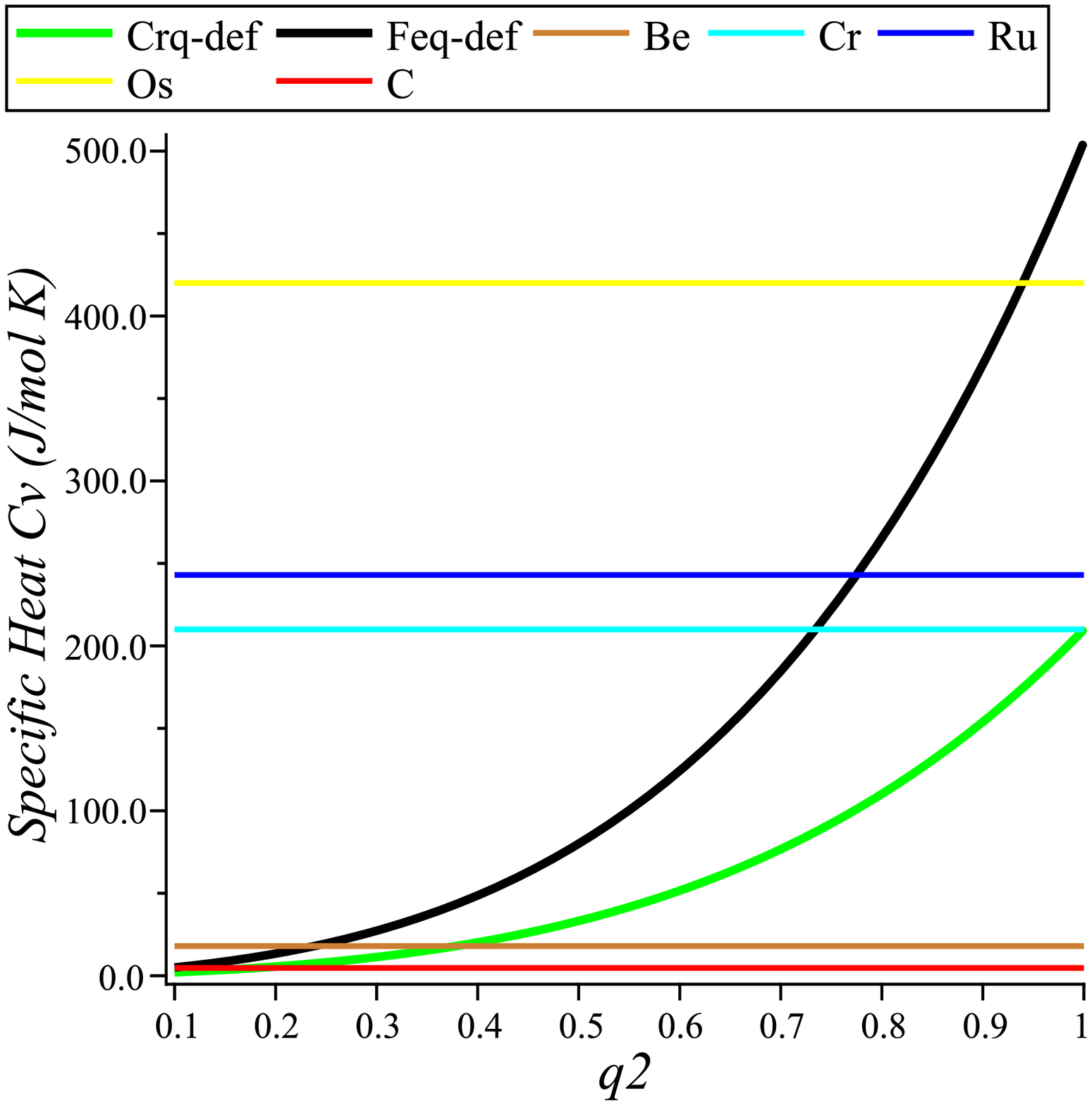}
}\caption{\small{Debye temperature $\theta_D$ depending on the variation $q_2=0.1,\cdots,1$ and $q_1=1$, 
\textbf{(left)}. Specific Heat $c_{V}$ depending on the variation $q_2=0.1,\cdots,1$ and $q_1=1$, \textbf{(right)}}}\label{gráficos 16}
\end{figure}

For illustration purposes, we choose iron (Fe) and chromium (Cr), two materials that can be employed in many areas of interest. 
In Figs.(\ref{gráficos 16}) we present deformed values of $Fe$ ($Fe_{q-def}$) 
(black curve) and $Cr$ ($Cr_{q-def}$) (green curve), for values $q_1=1$ and $q_2=0.1,\cdots,1$, where
for this range we assume the maximum deformation ($q_1=1$ and $q_2=0.1$) and the pure element 
 (bulk) ($q_1=1$ and $q_2=1$). The other elements are represented by colors and indicated in the very figure.

On the left side of the Fig.(\ref{gráficos 16}), we can observe that before reaching their limits, black and green 
curves can assume the values of Debye temperatures ($\theta_D$) of other elements. The $Fe_{q-def}$ e.g., 
equates to: beryllium (Be) when $q_2\approx 0.23$, chromium (bulk) (Cr) $q_2\approx 0.75$ and osmium (Os) $q_2\approx 0.94$. 
On the right, we have the behavior of the curves obtained for the specific heat $c_V$. We note that the behavior is quite 
different from the previous curves $\theta_D$, i.e., the curves start at lower values (maximum deformation) until they reach 
their pure values. Having $Fe_{q-def}$ as an example again, it is possible to see, 
as it reaches the value of specific heat capacity of all the elements, including $Cr$ (bulk) when $q_2\approx 0.74$.

\begin{figure}[hb]
\centerline{
\includegraphics[{angle=90,height=7.0cm,angle=270,width=7.0cm}]{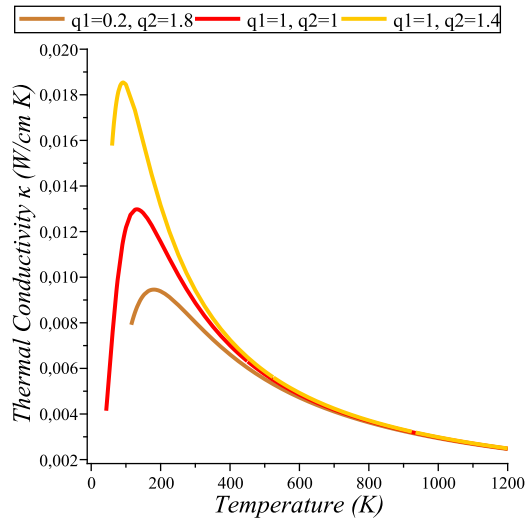}
}\caption{\small{Thermal conductivity $\kappa$ of $Fe$ as a function of temperature in the range of $T=0,\cdots,1200 K$, 
and  values for $q_1$ and $q_2$ given explicitly.}}\label{gráficos 20}
\end{figure}

Let us now return to Eq.(\ref{eq12.5}), where we have the complete Debye function $D(\alpha_{0_{q_1,q_2}})$ to see the behavior of the thermal conductivity. 
Thus, in the Fig.(\ref{gráficos 20}), we show a comparison to the thermal conductivity $\kappa$ for a pure and impure material.
We have the thermal conductivity as a function of temperature (T) for $Fe$ (bulk) and a combination of the values $q_1$ and $q_2$ for the $Fe_{q-def}$ (impure). 
Therefore, we have the $Fe$ (red) when $q_1=q_2=1$ (pure), and when $q_1=0.2$ and $q_2=1.8$, we have a similar behavior to silicon 
(Si) (golden), whose value for the Debye temperature is $\theta_D = 645 K$. Finally, for a combination of values $q_1=1$ and $q_2=1.4$ 
we have a curve that is similar to that of zinc (Zn) (orange), with $\theta_D = 327 K$. 

One should note that deformation is clearly playing the role of impurity concentration in the material sample. This is because 
deformation acts directly on the Debye temperature, which means that the Debye frequency is modified. Changing the Debye 
frequency is a clear sign of the material being modified by impurities.
\section{Conclusions}

The initial idea that $q$-algebra acts as a factor of disorder or impurity is enhanced by inserting two factors, 
the so-called Fibonacci oscillators. In Figs.(\ref{gráficos 16},\ref{gráficos 20}), we note that the elements that 
suffer deformation may become similar to others. 
The existence of more degrees of freedom as in the present case of two deformation parameters, $q_1$ and $q_2$, can be well associated with different types of deformations related to
two distinct phenomena of disorders or impurities such as, for instance, one due to pressure generating disorders and other due to doping, respectively.


\section*{Acknowledgments}

We would like to thank CNPq, CAPES, and PNPD/PROCAD-CAPES, for partial financial support.

\end{document}